\documentstyle[] {laa}
\begin{document}
\input epsf


\def\la{\hbox{\rlap{\raise.3ex\hbox{$<$}}\lower.8ex\hbox{$\sim$}\ }}
\def\ga{\hbox{\rlap{\raise.3ex\hbox{$>$}}\lower.8ex\hbox{$\sim$}\ }}


\def\mdot{$\dot{m}$}
\def\Lts{\langle L_{p,35} \rangle} 
\def\Ariel5{{\it Ariel 5}}
\def\uhu{{\it Uhuru}}
\def\la{\hbox{\rlap{\raise.3ex\hbox{$<$}}\lower.8ex\hbox{$\sim$}\ }}
\def\ga{\hbox{\rlap{\raise.3ex\hbox{$>$}}\lower.8ex\hbox{$\sim$}\ }}
\def\ctss{cts s$^{-1}$}
\def\Ltsu{{\langle L_{p,35} \rangle}^{-1}}
\def\etahx{\langle \eta_{\rm 40-200 keV} \rangle}
\def\etahxu{{\langle \eta_{\rm 40-200 keV} \rangle}^{-1}}
\def\ergs{ergs s$^{-1}$}
\def\ergscm{ergs s$^{-1}$ cm$^{-2}$}
\def\pho{ photons cm$^{-2}$ s$^{-1}$ }
\def\phokev{ photons cm$^{-2}$ s$^{-1}$ keV$^{-1}$}
\def\ap{$\approx$ }
\def\msol{$\rm M_\odot$ }
\def\ep{$\rm e^\pm$ }
\def\nh{H atoms cm$^{-2}$}
\def\fx{Flux$_{\rm 1-20~keV}$}
\def\fsoft{F$_{\rm 1-20~keV}$}
\def\fhard{F$_{\rm 20-200~keV}$}
\def\dmn{$\times 10^{-9}$ ergs s$^{-1}$ cm$^{-2}$}
\def\dmd{$\times 10^{-10}$ ergs s$^{-1}$ cm$^{-2}$}
\def\expa{$^{\rm a)}$}
\def\expb{$^{\rm b)}$}
\def\expc{$^{\rm c)}$}
\def\expd{$^{\rm d)}$}
\def\expe{$^{\rm e)}$}
\def\expf{$^{\rm f)}$}
\def\expg{$^{\rm g)}$}
\def\arcpt{$''\mkern-7.0mu .\mkern1.4mu$}
\def\ni{\noindent}
\def\na{\noalign}
\def\mc{\multicolumn}
\def\pb{\parbox}
\def\pb5{\parbox[t]{4.5cm}}
\def\erg{${\rm erg\;cm^{-2}\;s^{-1}}$}
\def\btsix{\beta_{rx}^{\rm 6\;keV}}
\def\bttwo{\beta_{rx}^{\rm 2\;keV}}
\def\mb{\makebox}
\def\dub#1{{\bf #1???}}
\def\sect#1{\section{#1}\typeout{#1}}
\def\percm{\ cm$^{-1}$}
\def\perccm{\ cm$^{-3}$}
\def\cm2sec{\ cm$^{2}$-sec}
\def\kms{km\ s$^{-1}$}
\def\micron{$\mu$m}
\def\microns{\ $\mu$m}
\def\arcsec{$\,^{\prime\prime}$}
\def\arcsecs{$\,^{\prime\prime}$}
\def\arcmin{$\,^\prime$~}
\def\arcmins{$\,^\prime$}
\def\deg{$^{\circ}$}
\def\ang{${\rm \AA}$}
\def\Halpha{H$\alpha$~}
\def\Hbeta{H$\beta$~}

%
\def\hr{\makebox[2.5mm][l]{$^{\rm h}$}}
\def\mi{\makebox[2.5mm][l]{$^{\rm m}$}}
\def\se{\makebox[0mm][l]{$^{\rm s}$}}
\def\dd{\makebox[2.5mm][l]{$^{\circ}$}}
\def\mm{\makebox[2.5mm][l]{$^{'}$}}
\def\ss{\makebox[0mm][l]{$^{''}$}}
\def\xx{\makebox[2.5mm][l]{ }}
\def\xs{\makebox[0mm][l]{ }}
\def\etal{{\it et al.\/}}
\def\FU{\ ergs cm$^{-2}$ s$^{-1}$}
\def\Htwo{H$_2$}
\def\H2S1{H$_2$ $S$(1)}
\def\Htwoline{\H2S1}
\def\HI{H\RomanI}       \def\HOne{H$^{+}$}
\def\HII{H\RomanII}
\def\HeI{He\RomanI}     \def\HeII{He\RomanII}
\def\Rsun{$R_\odot$}
\def\Lsun{$L_\odot$}
\def\Msun{$M_\odot$}
\def\aro{$\alpha_{ro}$}
\def\aox{$\alpha_{ox}$}
\def\aix{$\alpha_{ix}$}
\def\arx{$\alpha_{rx}$}
\def\bro{$\beta_{ro}$}
\def\btox{$\beta_{ox}$}
\def\brx{$\beta_{rx}$}
\def\airas{$\alpha(60/25)$}
\def\airass{$\alpha(100/60)$}
%
%
\def\etal{{\it et al.}}
\def\annrev{{\it Ann.\ Rev.\ Astron.\ Ap.}}
\def\aplet{{\it Ap.\ Letters}}
\def\aj{{\it Astron.\ J.}}
\def\apj{Ap J}
\def\apjl{{\it Ap.J. Letters}}
\def\apjlet{{\it Ap.\ J.\ (Lett.)}}
\def\apjs{{\it Ap.\ J.\ Suppl.}}
\def\apjsup{{\it Ap.\ J.\ Suppl.}}
\def\aasup{{\it Astron.\ Astrophys.\ Suppl.}}
\def\astro{{\it Astron.\ Astrophys.}}
\def\aa{{\it Astron.\ Astrophys.}}
\def\mnras{{\it M.\ N.\ R.\ A.\ S.}}
\def\nature{{\it Nature}}
\def\pasa{{\it Proc.\ Astr.\ Soc.\ Aust.}}
\def\pasp{{\it P.\ A.\ S.\ P.}}
\def\pasj{{\it PASJ}}
\def\pre{{\it Preprint}}
\def\qjras{{\it Quart.\ J.\ R.\ A.\ S.}}
\def\rppp{{\it Rep.\ ProAg.\ Phys.}}
\def\sovlet{{\it Sov. Astron. Lett.}}
\def\adspr{{\it Adv. Space. Res.}}
\def\expas{{\it Experimental Astron.}}
\def\ssr{{\it Space Sci. Rev.}}
\def\inpress{in press.}
\def\souspresse{sous presse.}
\def\inprep{in preparation.}
\def\enprep{en pr\'eparation.}
\def\submit{submitted.}
\def\soumis{soumis.}

\def\ergs{ergs s$^{-1}$}
\def\ergscm{ergs s$^{-1}$ cm$^{-2}$}
\def\pho{ photons cm$^{-2}$ s$^{-1}$ }
\def\phokev{ photons cm$^{-2}$ s$^{-1}$ keV$^{-1}$}
\def\ap{$\approx$ }
\def\msol{$\rm M_\odot$ }
\def\ep{$\rm e^\pm$ }
\newcommand{\vp}{\,{ v_{p}}}
\newcommand{\te}{\,{ T_{e}}}
\newcommand{\y}{\,{ y}}
\newcommand{\hzero}{\,{ H_{0}}}
\newcommand{\qzero}{\,{ q_{0}}}
\newcommand{\Inu}{\,{ I_{\nu}}}
\newcommand{\tr}{\,{ T_{r}}}
\newcommand{\mecdeux}{\,{ m_{e}c^{2}}}
\newcommand{\cd}{\,{ C_{d}}}
\newcommand{\td}{\,{ T_{d}}}
\newcommand{\neu}{\,{ n_{e}}}

    \thesaurus{Sect. 2         
              ( 12.03.1;  
                11.03.1; 
                11.09.3 
                             )
               }
   \title{Determination of the hot intracluster gas temperature
   		from submillimeter measurements}
 
   \subtitle{}

   \author{E. Pointecouteau 	,
          M. Giard 	 and
          D. Barret 	
          }
 
   \offprints{E. Pointecouteau, pointeco@cesr.fr}
 
   \institute{
Centre d'Etude Spatiale des Rayonnements, 9 avenue du Colonel Roche, BP-4346
       31029 Toulouse, France.}

   \date{Received               ; accepted              }
 
   \maketitle
 
   \begin{abstract}

Measurements of the Sunyaev-Zeldovich (hereafter SZ) distorsion of the cosmic microwave 
background
can give interesting physical informations on clusters of galaxies,
provided that the electronic temperature of the gas is known. Previous attempts
to do so have used the electronic temperature determination obtained from the 
X-ray spectra.
However, if the intergalactic gas is not homogeneous, the X-ray emission 
will trace the denser component, and the temperature
determination may not be relevant for the lower density gas which is dominating 
the SZ measurements. Moreover, the X-ray brightness decreases very rapidly with
the distance, which is not the case for the SZ effect.
Distant clusters might be detected 
from SZ measurements, whereas they are inaccessible
to X-ray observations. For these reasons, we have 
investigated the possibility to derive the electronic temperature of the gas
from the SZ measurements in the submillimeter range 
($\lambda \sim 300-600 \mu$m).
We show that given the sensitivities of the future
submillimeter space missions Planck Surveyor and FIRST, the electronic
temperature of massive clusters ($Y_{center}=3 \times 10^{-4}$) can be determined
with an accuracy ranging from 1 to 4 keV depending on its distance and the data available.

   \end{abstract}
 
%
\section{Introduction}

 The intergalactic medium is a strong source of diffuse X-ray radiation
by free-free emission (\cite{jones84}).
It is observed at submillimeter wavelengths too, via the Sunyaev-Zeldovich
(hereafter SZ) effect:
a spectral distorsion of the Cosmic Microwave Background (hereafter CMB) due to 
the interaction of
the electrons of the hot ionized gas with the photons of the CMB 
(\cite{Zel'dovich69}, \cite{sunyaev72}).
If the electronic temperature is determined from X-ray data, together with a 
model of the gas
distribution, SZ data allow to derive the Compton optical depth ($\tau$)  of 
the intergalactic gas.
This parameter directly provides the gas mass, if it is integrated over solid
angles. 
The association of X-ray and SZ data allows to estimate the Hubble constant, 
$\hzero$, independently
of the usual standard candles methods (see \cite{holzapfel97} for instance).
The peculiar velocity of several clusters can also be derived from the
Doppler effect, so that it should be
possible to detect the large scale gravitational field which is produced by 
the dark matter.

An important feature of the SZ brightness is that it is an absorption
effect on the CMB, which intensity is independent of z (if no evolution of the
cluster is assumed).
On the contrary, the cluster's X-ray surface brightness decreases with respect
to the usual $(1+z)^{-3}$ expansion factor and with respect to an  additionnal 
exponential factor, $exp(-E(1+z)/k\te)$, which becomes important for very distant
clusters.

In this paper, we investigate how the gas cluster temperature can be recovered from the SZ measurements
themselves.
In section 2, we present a simple Monte-Carlo method which allows to obtain the 
exact shape of SZ
spectra, taking into account the temperature dependence which is ignored in the 
usual analytical expression.
We emphasize the spectral effect due to the gas temperature which shows up
in the submillimeter domain.
In section 3 we quantify the error on the determination of the gas temperature  using SZ
measurements only, considering limitations of the instruments sensitivities and of the various
foreground and background emissions.

\section{Exact calculations of the SZ effect}

The frequency dependency of the SZ effect is usually approximated by a solution of the Kompaneets
equation (\cite{kompaneets57}) which is a second order approximation of the
Boltzmann equation (\cite{Rybicki79}). This solution needs:

\begin{equation}
\left. \frac{\Delta\Inu}{\Inu} \right |_{th}=\y f(x)
\end{equation}
for the thermal effect,
\begin{equation}
\left. \frac{\Delta\Inu}{\Inu} \right |_{cin}=-\frac{\vp \tau}{c} a(x)
\end{equation}
for the kinetic effect,\\
$f(x)$ and $a(x)$ are analytic functions of the dimensionless frequency:
\begin{equation}
x=\frac{h\nu}{k\tr}
\end{equation}
$y$ is the Comptonization parameter of the cluster: 
\begin{equation}
\y=\frac{k\te}{\mecdeux}\tau
\end{equation}

$\te$ is the electronic temperature of the intergalactic gas, $\vp$
is the peculiar velocity of the cluster, $\tau$ is the Compton optical
depth of the intergalactic gas. $T_{r}=2.726$K is the CMB temperature.
 
Photons of the CMB are scattered from low frequencies to higher frequencies.
The cross-over frequency is around $\nu=217$ GHz (e.g. $\lambda=1.38$ mm).

In case of millimeter and submillimeter SZ data, the analytic approximation
yield to analysis errors. Indeed, the electronic temperature level
implies weakly relativistic velocities for the electrons.
A few authors have worked on the relativistic corrections of the SZ effect
(\cite{wright79}, \cite{fabbri81}). 
Most recently, Rephaeli (\cite{rephaeli95}) has compiled
previous works to develop a semi-analytical treatment of the SZ effect and  
Stebbins (\cite{stebbins97}), Challinor \& Lassenby (\cite{challinor97}) and
Itoh et al. (\cite{itoh97})
have extended analytically the Kompaneets equation.
However, the exact SZ spectra can be obtained using a simple Monte-Carlo method 
which 
numerically integrates the transfer equation:
\begin{center}
\begin{equation}
\frac{\partial \Inu}{\partial s}=\neu\int d\beta \int d\Omega \; p_{e}(\beta) 
\frac{d\sigma}{d\Omega}
[\Inu(\nu_{1})-\Inu(\nu))]
\end{equation}
\end{center}

where $p_{e}(\beta)$ is the velocity distribution of the electrons,
      $\beta=$\large$\frac{v}{c}$ \normalsize,
      \large $\frac{d\sigma}{d\Omega}$ \normalsize is the differential 
scattering cross section,
      $\Inu$ is the intensity of the radiation at the frequency $\nu$ (before 
scattering) and $\nu_{1}$
      (after scattering),
      $\neu$ is the density of the electrons. 
      
The frequency shift of the photons is given by:

\begin{center}
\begin{equation}
\frac{\nu_{1}}{\nu}=\frac{1-B}{1+B_{1}+\frac{h\nu}{\mecdeux} (1-\cos \alpha)}
\end{equation}
\end{center}
where $B=\beta \cos \theta$ and $B_{1}=\beta \cos \theta_{1}$ and
$\cos\alpha=\cos\theta\cos\theta_{1}+\sin\theta\sin\theta_{1}\cos(\phi-\phi_{1})$,
($\theta$,$\phi$) and ($\theta_{1}$,$\phi_{1}$) are the angles between the
electron's and the photon's directions of propagation,
respectively before and after the scattering.

The exact differential cross-section has to be used (\cite{podzniakov83}):
\begin{equation}
\frac{d\sigma}{d\Omega}=\frac{r_{e}^{2}}{2}\frac{1-B}{\gamma^{2}(1+B_{1})^{2}}
(1+[1-\frac{1-\cos\alpha}{\gamma^{2}(1-B)(1-B_{1})}]^{2})
\end{equation}
$r_{e}=e^{2}/\mecdeux$ is the classical radius of the electron and 
$\gamma=(1+\beta^{2})^{-1}$
is the Lorentz's factor.

\begin{figure}[h]
\vbox to 6cm{
\epsfxsize=8.5cm 
\epsfbox{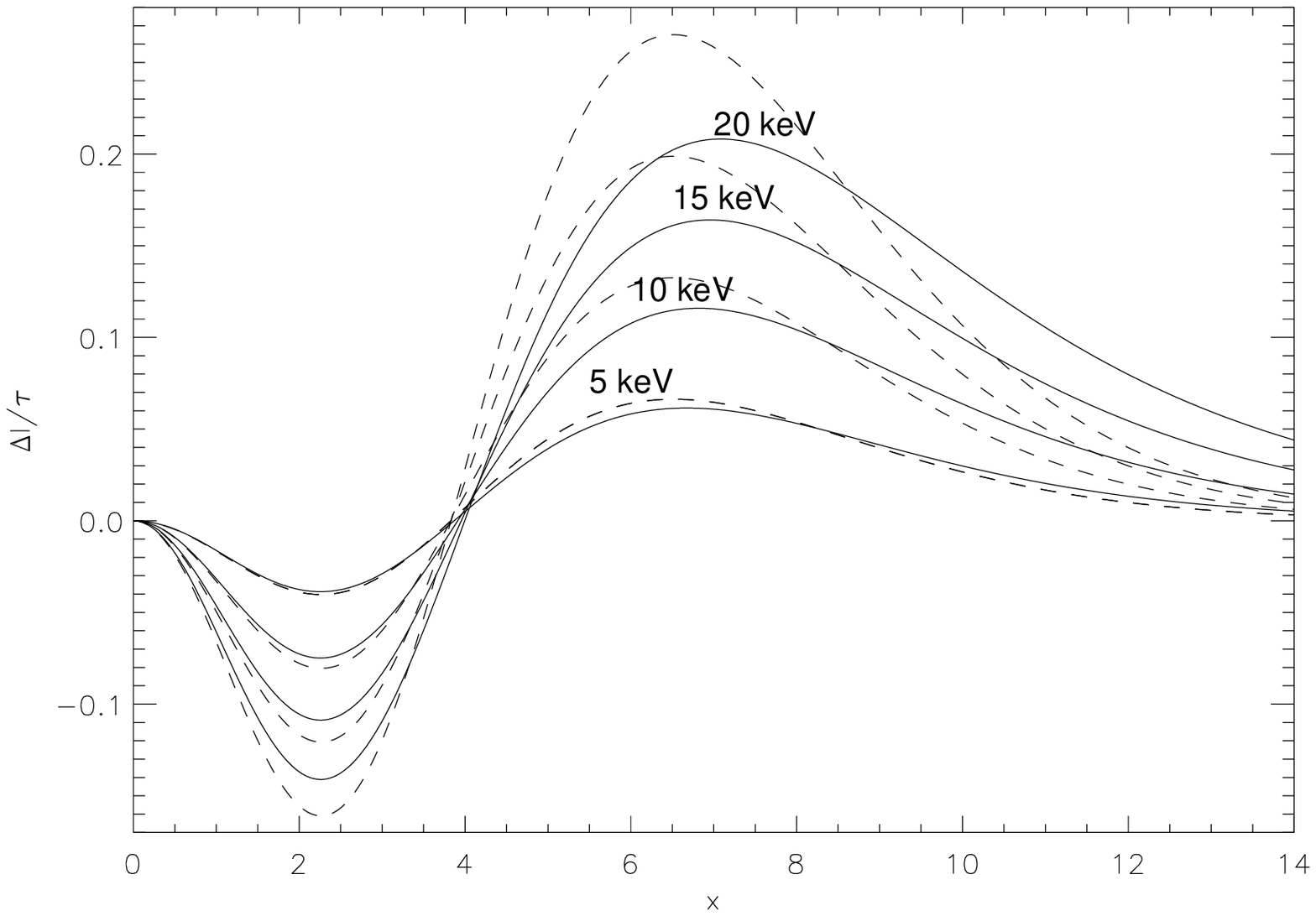}
}
\caption[]{\label{Fig. 1}
\normalsize
Comparaison of the SZ spectra obtained using the analytical approximation 
(dotted lines) and those obtained using the exact Monte Carlo calculation (solid lines).
Spectra are plotted for clusters with 5, 10, 15 and 20 keV temperature
(in units $(hc)^{2}/[2(kT_{e})^{3}]$).

}
\end{figure}

As shown on Fig. \ref{Fig. 1}, the spectra that we obtained are in accordance with those 
calculated by Rephaeli
(\cite{rephaeli95}).
Those spectra can be obtained from our anonymous ftp site
(ftp.cesr.fr/pub/astrophysique/sz/).
We will show hereafter that the SZ dependency on $\te$ can be used to derive
the intracluster gas temperature from submillimeter data (see Fig. 1).

\section{\label{model} Determinations of the electronic temperature}

\begin{table*}[t]
\begin{center}
\begin{tabular}{l|cccc|ccccccc}
\hline
&&&FIRST&&&&&&PLANCK \\
\hline
  $\nu\; _{(GHz)}$ & 1765& 1200& 857& 545  &857& 545& 353& 217& 150& 100 \\ 
  $\theta\;_{(arcmin)}$ & 4.5$^{(1)}$& 4.5$^{(1)}$& 4.5$^{(1)}$& 4.5$^{(1)}$&  4.5& 4.5& 4.5& 4.9& 7.4& 10.6 \\
  $NEB\;_{(10^{-2}\; MJy/sr/\surd Hz)}$ & 11.0$^{(1)}$& 8.1$^{(1)}$& 6.2$^{(1)}$& 4.3$^{(1)}$&  6.1& 5.9& 2.9& 2.6& 1.4& 1.0 \\
\hline
  rms $I_{DUST}\;_{(10^{-2}\; MJy/sr)}$ & 63.& 44.& 22.& 6.4&  22.& 6.4& 1.5& 0.27& 0.06& 0.01 \\
  rms $I_{ULIRG}\;_{(10^{-2}\; MJy/sr)}$ & 0.61& 0.59& 0.45& 0.021&  4.8& 2.2& 0.7& 0.12& 0.01& 0 \\
  rms $I_{CMB}\;_{(10^{-2}\; MJy/sr)}$ & 0& 0& 0.01& 0.47&  0.01& 0.47& 2.5& 3.9& 3.1& 1.9 \\
  rms $I_{ff+synch}\;_{(10^{-2}\; MJy/sr)}$ &  0.034& 0.039& 0.041& 0.046& 0.041&  0.046& 0.051& 0.058& 0.066& 0.074 \\
  $f_{dill}(z=0.1)$ &  0.87& 0.87& 0.87& 0.87& 0.87&  0.87& 0.87& 0.85& 0.74& 0.62 \\
  $f_{dill}(z=1)$ &  0.43& 0.43& 0.43& 0.43& 0.43&  0.43& 0.43& 0.41& 0.28& 0.2 \\
\hline
\end{tabular}
\end{center}
\caption{\label{table1} Characteristics of the Planck Surveyor's and the FIRST's passbands
used. The Noise Equivalent Brightness (e.g. NEB) is given
for the nominal beams for Planck and for a virtual beam for FIRST $^{(1)}$.
The rms contribution of the backgrounds and foregrounds are given for 
the beam used (see caption of Fig. 1 and text). Dilution factors applied to
$Y_{center}=3\times 10^{-4}$ are given for redshifts $z=0.1$ and $z=1$
and for a cluster with  $R_{c}=0.3$ Mpc and $\Omega _{tot}=0.3, H_{0}=70$ km/Mpc/s.}
$^{(1)}$ A virtual beam of 4.5 arcmin (e.g. the Planck submillimeter beam) is used
in simulations.
It will be obtained by summation of adjacent sky pixels. The NEB in each bands is
the sum of photon noise and detector noise, as determined in section 3.2.1,
modified by a factor equal to the ratio of the diffraction limit beam to virtual
beam diameters

\end{table*}

In a few years the Planck Surveyor and FIRST missions will detect the SZ effect
with a very good sensitivity. 
It is likely that new distant clusters will be identified with SZ data and that these clusters will
remain out of reach for the X-ray and/or optical telescopes.
In the following, we discuss the two ways of determining the electron
temperature for galaxy clusters from X-ray or SZ data.

\subsection{\label{c/s} Precision of X-ray determinations}

Using the XSPEC software (\cite{arnaud96}) we simulated a 25
kilosecond ASCA (GIS2) observation of a galaxy cluster with the
following spectral parameters: Bremsstrahlung temperature ($\te$) of 8
keV, an unabsorbed 1-10 keV flux of $6 \times 10^{-11}$ \ergscm~
corresponding to a rich cluster with $0<z<0.2$ observed through a
column density of $10^{21}$\nh. For such an observation, $\te$ is
recovered with an error of typically 0.5 keV at 68\% confidence
level. These errors are consistent with those derived from actual
observations (David \cite{David93}). Given the ASCA energy range
(0.1-10 keV), the error on $\te$ should increase with $\te$.
This is simply because as $\te$
increases the cutoff in the Bremsstrahlung spectrum moves at higher
energies, and approaches or even exceeds the high energy threshold of
the instrument where its sensitivity drops sharply. Consequently,
fitting the spectrum will tend to underestimate $\te$. This is a major
limiting factor which is primarily related to the energy coverage of
the instrument. This will also apply to future instruments, like XMM,
although their sensitivities are much better than ASCA.  For
instance simulating an ASCA observation with the same input spectrum as
above but with $\te=12$ keV, the fit recovers $\te$ as $10.9 \pm 0.9$ keV
(90\% confidence) barely consistent with the input value. Obviously
the recovered $\te$ and associated errors via X-ray observations depends
also on the input flux which has to be compared with the instrument
detection sensitivity. As the X-ray flux at earth decreases sensitively
with the redshift ( e.g. $z\ge 1$), for distant clusters the accuracy in
determining $\te$ will also decrease.
For instance, Hattori et al. (\cite{hattori97}) fitted the ASCA data of the
AXJ2019+1127 X-ray cluster to $z=0.94$ and $\te=8.6^{+4.2}_{-3.0}$K.

In the case of an inhomogeneous intracluster gas, if we considered 
the temperature-density correlation, a $\te$ determination via
X-ray emission ($\propto \int{\neu^{2}(l) dl}$) may miss a low density
component of the gas, so that an independent temperature determination,
via SZ measurements (mostly sensitive to a low component, e.g.
$\propto \int{\neu(l)dl}$) is also interesting.

\subsection{\label{c/s} Temperature determination with Planck and FIRST}

In the following, we have simulated SZ measurements of a  rich cluster performed
with Planck and FIRST. The photometric bands used are those of the HFI for Planck,
PHOC and SPIRE for FIRST.  Their characteristics are summarized in Table
\ref{table1}. They are taken from the COBRAS/SAMBA Phase A report concerning
Planck (except for the additional bolometer channel at 100 GHz). Concerning FIRST, we
have assumed that the noise level used is the quadratic sum of the photon noise, 
$NEP_{phot}=h\nu (\frac{S \Omega}{\lambda ^{2}}\, \Delta \nu \, \epsilon \eta
n(1+\epsilon \eta n))^{1/2}$ and the detector noise $NEP_{det}=3\times 10^{-17}$ 
W/$\surd$Hz. $S \Omega /\lambda ^{2}=1$
at the diffraction limit, $\Delta \nu /\nu \simeq 0.3$, $\epsilon =0.03$ is the 
telescope emissivity, $\eta =0.5$ is the system response (including
transmission and detector efficiency),
$n=2/(e^{(\frac{h \nu}{kT_{tel}})}-1)$ is the photon phase space occupation 
number for the telescope thermal emission, $T_{tel}=80$ K (satellite at $L_{2}$
Sun-Earth Lagrangian point.). The wavelengths of the FIRST 
lower frequency channels have been adjusted to the Planck higher frequency bands.

We estimate the precision of the temperature determination by repetitive least square
fits on simulated 
submillimeter and millimeter data: Planck alone or Planck plus FIRST combination.
We take into account both the instrumental noise and the contaminating sky emissions:
galactic dust, free-free, synchrotron, Ultra-Luminous InfraRed Galaxies (ULIRGs),
and CMB. We did not attempt to determine the velocity through
the kinetic effect since it is spectraly identical to the primary CMB and has already
been extensively studied (see \cite{haehnelt96} and \cite{aghanim97}).

\subsubsection{\label{signal} Simulation of clusters observations}

We simulate a massive cluster, $Y_{center} = 3 \times 10^{-4}$, with isothermal 
$\beta$ density gas profile, $T_e = 8$keV, $\beta = 2/3$, $R_{c} = 0.3$ Mpc,
$n_{0} = 1.9\times 10^{-2} cm^{-3}$. We take into account the dilution of the cluster SZ
emission in the instrumental beam by proper integration of the SZ profile
over the beam (we assume the density cutoff at 15 $R_{c}$). Concerning the
FIRST channels we use with a virtual beam equal to the Planck submillimeter
beam ($4.5$ arcmin) which will be obtained by the summation of adjacent sky
pixels. The surface brightness noise level (NEB) is correspondingly improved
from the diffraction limit value by a factor equal to the ratio of the
diffraction limit beam to the virtual beam.
The signal is then correctly integrated over the finite bandpass of the instruments.

We then add the following astrophysical contributions:

- A dust signal with rms level $I_{100 \mu} = 0.3$ MJy/sr,  value which is not
exceeded on 19$\%$ of the whole sky(This percentage is determined from
an all sky map of the rms 100 $\mu$m fluctuations calculated in bins of
0.7$^{\circ}\times$0.7$^{\circ}$ from the ISSA IRAS maps). A spectral
characteristics $T = 17.5 K$ , $n = 2$, typical of high latitude clouds
(Boulanger et al. \cite{boulanger96}). The variability  of the dust spectrum
is taken into account by introducing
a random fluctuation in the spectral index. This is fixed to 10$\%$
rms of the average value ($n=2$), in accordance with measurements of
the 2 meter submillimeter stratospheric telescope PRONAOS on high
latitude galactic cirrus (Bernard et al., in preparation).

- Fluctuations of the integrated contribution from the background ULIRGs. Their
rms level is integrated from the number counts of Guiderdoni et al. \cite{guiderdoni97}.
The flux limit is set such that the probability to find a source brighter than that
limit within the beam is smaller than 10$\%$ for the Planck's bands. For FIRST,
considering the high angular resolution available, we assume that all sources above
the 3$\sigma$ noise level are detected and subtracted before summation in
the 4.5 arcmin virtual beam.

- Fluctuations of the primary CMB at a rms level of $\Delta T/T = 30\times 10^{-6}$
(e.g. $\Delta \vp \simeq 500$ km/s for our cluster). 

- free-free and synchrotron: Their level and spectral
behaviour are taken from the COBE determination (Bennett et al. \cite{bennett92}):
$7 \mu K$ rms free-free at 53 GHz with spectral index n = -0.16, 
$4 \mu K$ rms synchrotron at 53 GHz with spectral index n = -0.9.
(The COBE measurement has been extrapolated to sub-degree angular scales
assuming $l^{-3}$, \cite{bersanelli96}).

A random instrumental noise with rms level as in Table \ref{table1} is then added
to the astrophysical signal. Finally, a $1 \%$ relative error is randomly
added to all bands to take into account the band to band calibration uncertainty.

\begin{figure}[t]
\vbox to 6cm{
\epsfxsize=8.5cm \epsfbox{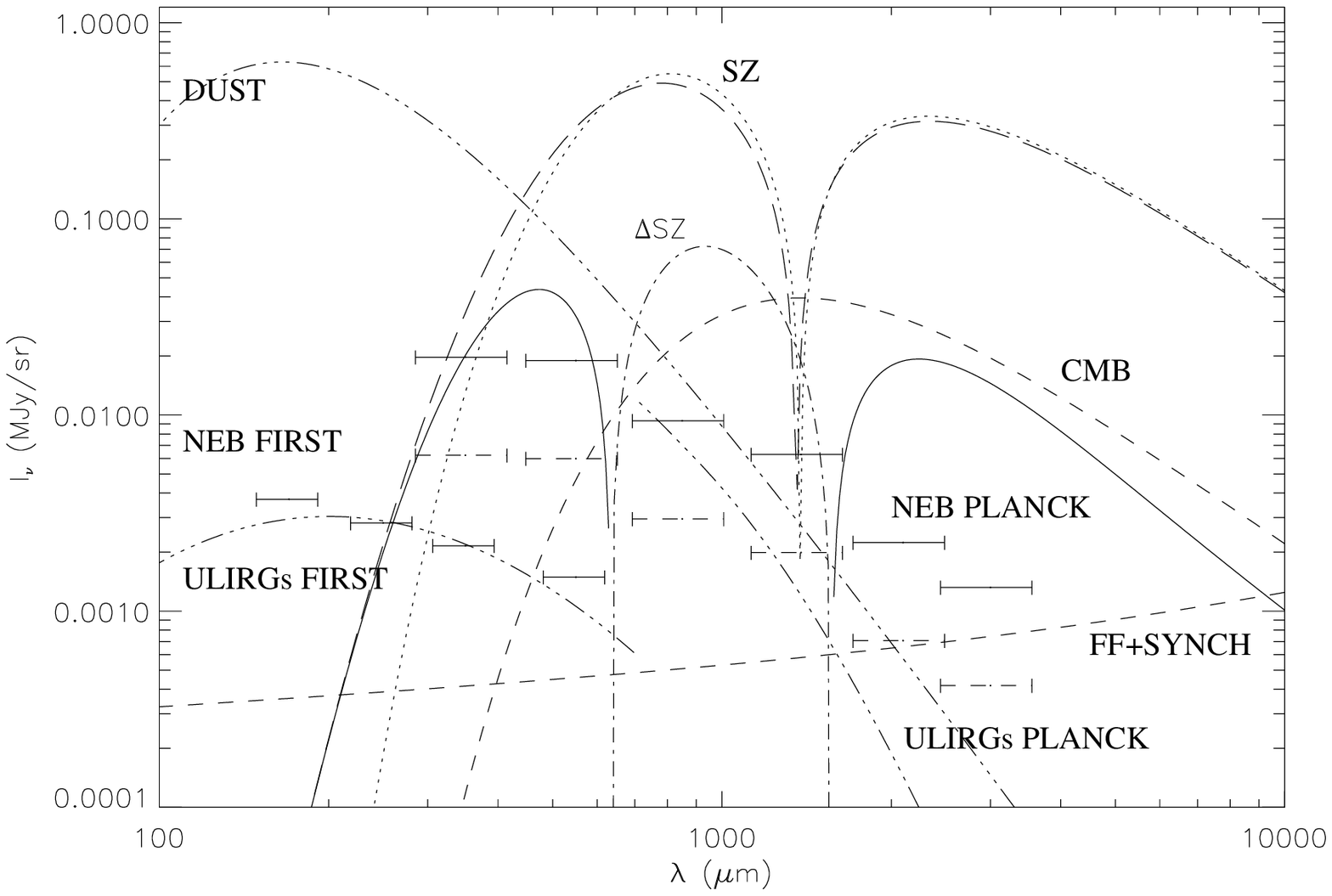}
}
\caption[]{\label{Fig. 2}
\normalsize
The temperature effect in the thermal SZ effect 
($\Delta SZ$=difference between exact and approximation calculation): 
solid line for the positive part and dotted-dashed line for the negative part.
The exact SZ effect (long dash line)
and the analytical approximation (dotted line) are overplotted.
($\te=8$ keV, $Y _{center}=3 \times 10^{-4}$).
Other contributions have been plotted: The CMB at the level of $\Delta T/T=30
\times 10^{-6}$, the galactic free-free plus synchrotron emission respectively
normalized at the level of $7 \mu$K and $4 \mu$K  at 53 GHz (dashed lines). 
The galactic dust (with $T_{d}=17.5$ K , $n=2$ and normalized to 0.3 MJy/sr at
100 $\mu$m) and the background infrared galaxies beam to beam fluctuations
(3 dot-dashed lines). (The background galaxies fluctuations are plotted as two 
curves, for the Planck and for the FIRST channels, as described in the text.). 
Horizontal bars represent the Plank and the FIRST sensitivities levels.
(Dashed bars for the high sensitivity regions of the Planck survey.)
}
\end{figure}

The different emissions and noise levels are shown in Fig. \ref{Fig. 2} together
with the instrumental band positions.

\subsubsection{\label{restore} Restoration of the cluster's parameters}

We fit a five parameters model to the simulated multi-band measurements. This model
is the sum of : SZ thermal effect ($\te, Y_{center}$), primary CMB 
($\Delta T/T$) and dust ($C_{d}, n$). $C_{d}$ and $n$ are respectively the level
and spectral index of the dust.
The model also take into account the integration over different spectral bands.

The errors on the restored parameters are estimated from the statistics of the
multiple simulations and fits. The ULIRGs are not explicitly included in the
separation because their spectrum is very close to galactic dust spectrum, 
however they are partly taken into account through the variable 
spectral index of the fitted dust.

\section{\label{first} Results and discussion}

Due to the likely curvature of the universe,
a massive cluster as the one considered in the previous section ($r_{c}>0.3$ 
Mpc), will never show a SZ profile having a width at half maximum less than 2
arcmin.
The integration time ($\sim 10$ s) is the main limiting factor
concerning the use of the Planck survey to measure the SZ effect with a high 
accuracy.
Actually, this should be overcome by the FIRST telescope.
Whatever are the missions final concepts and observing strategies, data from the
two satellites will be available in a few years, and will provide a
unique tool to study distant clusters of galaxies via precise 
measurements of the submillimeter SZ effect.

The accuracy on the determination of $\te$ is strongly dependent
on the integration time.
The FIRST Deep survey, a foreseen key program (\cite{beckwith93}), is
expected to cover 100 square
degrees on the sky, for a total time of $2\times 10^{6}$ seconds that corresponds
to an integration time of 500s per position, which is the value that we will use
in the following.
According to the Planck's mission observing strategy, some parts of the sky
will be observed with an integration time a factor of 10 higher than the nominal
time. 
(about 100 squared degrees on the sky.).

We restore the gas temperature for these two integration times in the
cases of Planck alone and the FIRST/Planck data combination.
For Planck alone, effective beams are all fixed to 10.6 arcmin (the 3mm beam of Planck)
to obtained a better sensitivity. 

Results are summarized in Table \ref{table 2}.\\

The accuracy on the estimation of the SZ parameters and the number of clusters 
that will be detectable depends on the observing strategy.
A first one would be to search for rich clusters in the FIRST deep survey fields,
completed by the corresponding positions of the Planck survey.
One can expect that part of the FIRST deep survey will be performed in a
region where Planck has a higher integration time per pixel (P2/F case in Table 2).
The precision on the temperature determination is in this case
of 1.2 keV ($z=0.1$) to 3.9 ($z=1.$).
Massive distant clusters can be identified in the FIRST data as positive extended
excesses at the larger wavelengths, after removal of the dust component.
(The dust contamination should not be a problem since the deep survey field will
be selected among the best galactic windows).

A second strategy will consist of a search for rich clusters in the Planck
survey, completed by pointed observations with FIRST.
The distant rich clusters will show almost no spatial extension in the Planck survey
($FWHM_{SZ}\simeq 3.5\theta _{c}\simeq 2.2$ arcmin at $z=1$) and will be recognized
among the faint clusters detected in the Planck survey, because they will have no
or only faint X-ray counterparts.
Actually, massive clusters (e.g: $y>3.\times 10^{-4}$) will be detectable in the
Planck survey data  at all redshifts: the most extreme 
dilution factor for such a cluster is about 0.18 (10.6 arcmin beam)
and the instrument sensitivity to SZ effect is of order $1.34\times 10^{-6}$
(no dilution).
For each cluster, depending on the detector array size, a five arcmin field 
on the cluster (plus comparison field) can 
be covered with FIRST within a few hours of observation.
The precision on the temperature determination will be of order 1.2 to 1.4 keV
for a cluster at $z=0.1$, depending on the Planck integration time on this cluster.
For very distant clusters the dilution is important and the precision is limited
to $\sim$4 keV due to the contamination of the infrared background galaxies. This
result, which is comparable to the precision of the current X-ray
determinations (Hattori et al. \cite{hattori97}), could be improved by cleaning properly 
the emission in the Planck bandpass from the infrared galaxies contamination.

Depending on the age of the universe, its geometry and the epoch of cluster
formation, there may be a few or a large number of distant rich clusters 
(\cite{oukbir97}). 
Although cosmological standard models (CDM and $\Omega \simeq 1$) favour 
the formation of rich structures only at a recent time, 
and seem to be supported by the X-ray cluster's distribution at $z<0.5$ 
(\cite{bartlett94}, \cite{luppino95}),
several detections of galaxies and structures have been reported
at very high z ($1 < z < 5$): \cite{pascarelle96}, \cite{malkan96}, \cite{lefevre96}
and \cite{dey98} at optical wavelengths,
the X-ray detection of AXJ2019+1127 by Hattori et al. (\cite{hattori97}) at z=0.94.
Most unexpected have been the detections by \cite{jones97} and \cite{Richards97} of 
negative decrements at centimeter wavelengths in the direction of distant quasars. 
These decrements can be
attributed to massive clusters at $z > 1$, unobservable at other wavelengths. 

Korolyov et al. (\cite{korolyov86}) were the first to write that it is easier to 
detect high redshift clusters of galaxies in microwave spectral band than in X-rays. 
If this is confirmed, it  means that a fairly large number of rich clusters may be 
detectable in the Planck survey and selected on the basis of low dust 
contamination and faint (or no) X ray 
counterparts for observations with FIRST.

\begin{table}[t]
\begin{center}
\begin{tabular}{lccccccc}
&&&$\Delta \te$ (keV) \\
\hline
$\Delta \te$ (keV) && P1 & P2 && P1/F & P2/F \\
\hline
z=0.1 && 3.1 & 2.2 && 1.4 & 1.2  \\
z=1. && - & 4.2 && 4.3 & 3.9\\
\hline
\end{tabular}
\end{center}
\caption{\label{table 2} 1 $\sigma$ statistical error (keV) on the determination of the gas
temperature (8 keV cluster) for different configurations. For Planck alone, P1: nominal integration time.
P2:10$\times$nominal integration time. For Planck/FIRST, P1/F: Planck's nominal
integration time and First deep survey (DS) for FIRST. P2/F: Planck 10$\times$nominal
time and First deep survey
}
\end{table}

\acknowledgements{}

We thank N. Aghanim, F.X. Desert and F. Bouchet for their very helpful suggestions.
We are very grateful to the referee, R. Sunyaev, for his fruitful comments.


\begin{thebibliography}{}

\bibitem[Aghanim et al. 1997]{aghanim97}
Aghanim, N., et al., 1997, A\&A, 325, 9

\bibitem[Arnaud 1996]{arnaud96}
Arnaud, K.A., 1996, Astron. Data
Analysis Software and Systems V, eds. Jacoby G. and Barnes J., p17, ASP
Conf. 101.

\bibitem[Bartlett $\&$ Silk 1994]{bartlett94}
Bartlett, J.G., Silk, J., 1994, ApJ 423, 12

\bibitem[Beckwith et al. 1993]{beckwith93}
Beckwith, S., et al., 1993, FIRST Red book, ESA report

\bibitem[1992]{bennett92}
Bennett, C., et al., 1992, ApJ, 391, 466

\bibitem[Bersanelli et al. 1996]{bersanelli96}
Bersanelli, M.et al., 1996, COBRA/SAMBA ESA report

\bibitem[1996]{boulanger96}
Boulanger, F., et al., 1996, A\&A, 312, 256

\bibitem[1997]{challinor97}
Challinor, A., Lasenby, A., 1997, astro-ph/9711161

\bibitem[1993]{David93}
David, L.P., et al., 1993, ApJ 412, 479

\bibitem[Dey et al. (1998)]{dey98}
Dey, A., et al., 1998, astro-ph/9803137

\bibitem[Fabbri 1981]{fabbri81}
Fabbri, R., 1981, A\&SS 77, 529

\bibitem[(1997)]{guiderdoni97}
Guiderdoni, B., et al., 1997, MNRAS, 295, 877

\bibitem[1997]{hattori97}
Hattori, M., et al., 1997, Nat, 388, 146

\bibitem[Haehnelt \& Tegmark 1996]{haehnelt96}
Haehnelt, M., G., Tegmark, M., 1996, MNRAS, 279,545

\bibitem[Holzapfel et al. 1997]{holzapfel97}
Holzapfel, W., L., et al. , 1997, ApJ, 480, 449

\bibitem[1997]{itoh97}
Itoh, N., Kohyama, Y., Nozawa, S., 1997, astro-ph/9712289

\bibitem[Jones $\&$ Forman 1984]{jones84}
Jones, C., Forman, W., 1984, ApJ 276,38

\bibitem[Jones et al. (1997)]{jones97}
Jones, M.E., et al., 1997, ApJ, 479, L1

\bibitem[Kompaneets 1972]{kompaneets57}
Kompaneets, A.S., 1957, Sov. Phys.-JETP, 4, 730

\bibitem[1986]{korolyov86}
Korolyov, V.A., Sunyaev, R.A., Yakubtsev, L.A., 1986, 
Sov. Astron. Letters, 12, 141

\bibitem[Le Fevre et al. (1996)]{lefevre96}
Le Fevre, et al., 1996, ApJ, 471, L11

\bibitem[Luppino \& Gioia 1995]{luppino95}
Luppino, G. A., Gioia, I., M.,1995, ApJ, 445, L77

\bibitem[Malkan et al. (1996)]{malkan96}
Malkan, M., A., Teplitz, H., McLean, I., S., 1996, ApJ, 468, L9

\bibitem[Oukbir \& Blanchard 1997]{oukbir97}
Oukbir, J., Blanchard, A., 1997, A\&A, 317, 1

\bibitem[Pascarelle et al. (1996)]{pascarelle96}
Pascarelle, S., et al., 1996, ApJ 456, L21
 
\bibitem[Podzniakov, Sobol \& Sunyaev 1983]{podzniakov83}
Podzniakov, L.A., Sobol, I.M. and Sunyaev, R.A., 1983, Sov. Scientific 
Review-A\&SS,
Sunyaev-Hardwood Academy Publishers, New York, vol 2, p 189

\bibitem[1995]{rephaeli95}
Rephaeli, Y., 1995, ApJ 445, 33

\bibitem[Rybicki \& Lightman 1979]{Rybicki79}
Rybicki, G.B., Lightman, A.P., 1979, ''Radiative Process in Astrophysics'',
New York: Wiley $\&$ Sons

\bibitem[Richards et al. (1997)]{Richards97}
Richards, E., et al, 1997, AJ 113, 1475

\bibitem[1997]{stebbins97}
Stebbins, A., 1997, astro-ph/9709065

\bibitem[Sunyaev \& Zeldovich 1972]{sunyaev72}
Sunyaev, R.A., Zeldovich, Ya.B., 1972, Comments Astrophys. Space Phys., 4, 173

\bibitem[Wright 1979]{wright79}
Wright, E.L., 1979, ApJ 232, 348
 
\bibitem[Zel'dovich \& Sunyaev 1969]{Zel'dovich69}
Zeldovich, Ya.B., Sunyaev, R.A., 1969, Ap\&SS 4, 301

\end{thebibliography}
\end{document}